\documentclass[%
 aip,
 amsmath,amssymb,
 reprint,
]{revtex4-1}

\usepackage{graphicx}
\usepackage{dcolumn}
\usepackage{bm}

\usepackage{verbatim} 
\usepackage[utf8]{inputenc}
\usepackage[T1]{fontenc}
\usepackage{mathptmx}

\usepackage{xcolor}

\usepackage{natbib}
\usepackage[colorlinks, citecolor=red]{hyperref}

\preprint{APS/123-QED}
\begin{document}

\title{Noise effects in the nonlinear thermoelectricity of a Josephson junction} 

\author{G. Marchegiani}%
 \email{giampiero.marchegiani@nano.cnr.it}
\affiliation{
NEST Istituto Nanoscienze-CNR and Scuola Normale Superiore, I-56127 Pisa, Italy}%
\author{A. Braggio}%
 \email{alessandro.braggio@nano.cnr.it}
\affiliation{
NEST Istituto Nanoscienze-CNR and Scuola Normale Superiore, I-56127 Pisa, Italy}%
\author{F. Giazotto}%
 \email{francesco.giazotto@sns.it}
\affiliation{
NEST Istituto Nanoscienze-CNR and Scuola Normale Superiore, I-56127 Pisa, Italy}

\date{\today}

\begin{abstract}
We investigate the noise current in a thermally biased tunnel junction between two superconductors with different zero-temperature gaps. When the Josephson effect is suppressed, this structure can support a nonlinear thermoelectric effect due to the spontaneous breaking of electron-hole symmetry, as we recently theoretically predicted. We discuss the possibly relevant role played by the noise in the junction. While a moderate noise contribution assists the generation of the thermoelectric signal, further unveiling the spontaneous nature of the electron-hole symmetry breaking, a large noise contribution can induce a switching between the two stationary thermoelectric values, thus hardening the detection of the effect and its application. We demonstrate that the thermoelectric effect is robust to the presence of noise for a wide range of parameters and that the spurious fluctuations of the thermoelectric signal can be lowered by increasing the capacitance of the junction, for instance by expanding the junction's size. Our results pave the way to the future experimental observation of the thermoelectric effect in superconducting junctions, and to improved performance in quantum circuits designed for thermal management.
\end{abstract}


\maketitle 

One of the relevant factors in the development of quantum technologies~\cite{Riedel_2017,Acn2018} is the control over thermal currents, since both the low-operating temperatures and the reduced size and dimensionality typically limit the heat flow~\cite{Pop2010,Liu2012}. This motivated an intense investigation of the thermal transport both on the theoretical and on the experimental sides~\cite{Goodson2006,GiazottoRMP2006,WangEPJB2008,Pop2010,DiVentraRMP2011,Yang2012,Muhonen2012,CahillReviewII2014,BergfieldMolecular}. Physical systems based on hybrid superconducting junctions~\cite{Tinkham2004,barone1982physics} have demonstrated a great potential for heat management issues. For instance, the electronic refrigeration~\cite{RajauriaPRL99,GiazottoRMP2006,Muhonen2012}, and the phase-coherent modulation of thermal currents~\cite{GiazottoNature,FornieriReview} are experimentally well established. More recently, they have also attracted interest for their good thermoelectric performance, when put in close proximity with topological materials~\cite{BlasiPRL} or ferromagnetic elements~\cite{BelzigTEPRL,Ozaeta2014,Kolenda2017,KolendaPRL,VirtanenReviewTE}, and may find use for the detection of radiation or thermometric applications~\cite{GiazottoThermometerNFIS}. In order to have an enhanced response, these thermally activated structures typically require a strong temperature gradient~\cite{GuarcelloPRAppliedMemory,MarchegianiEngine,MarchegianiCooler}, and the mean thermal energy comparable with the relevant energy scale, such as the superconducting gap. In this condition, it is essential to quantify the effect of the noise in the system, which may be crucial for a successful experimental implementation.

In this letter, we investigate the impact of the noise on the nonlinear thermoelectric effect in superconducting junctions predicted in Ref.~\onlinecite{MarchegianiPRL}, where we discussed how thermoelectricity can arise even in the presence of electron-hole symmetry, due to a spontaneous symmetry-breaking. We demonstrate that the effect is robust against noise and amenable to experimental observation, for a wide range of parameters.

\begin{figure}[ht]
	\begin{centering}
		\includegraphics[width=\columnwidth]{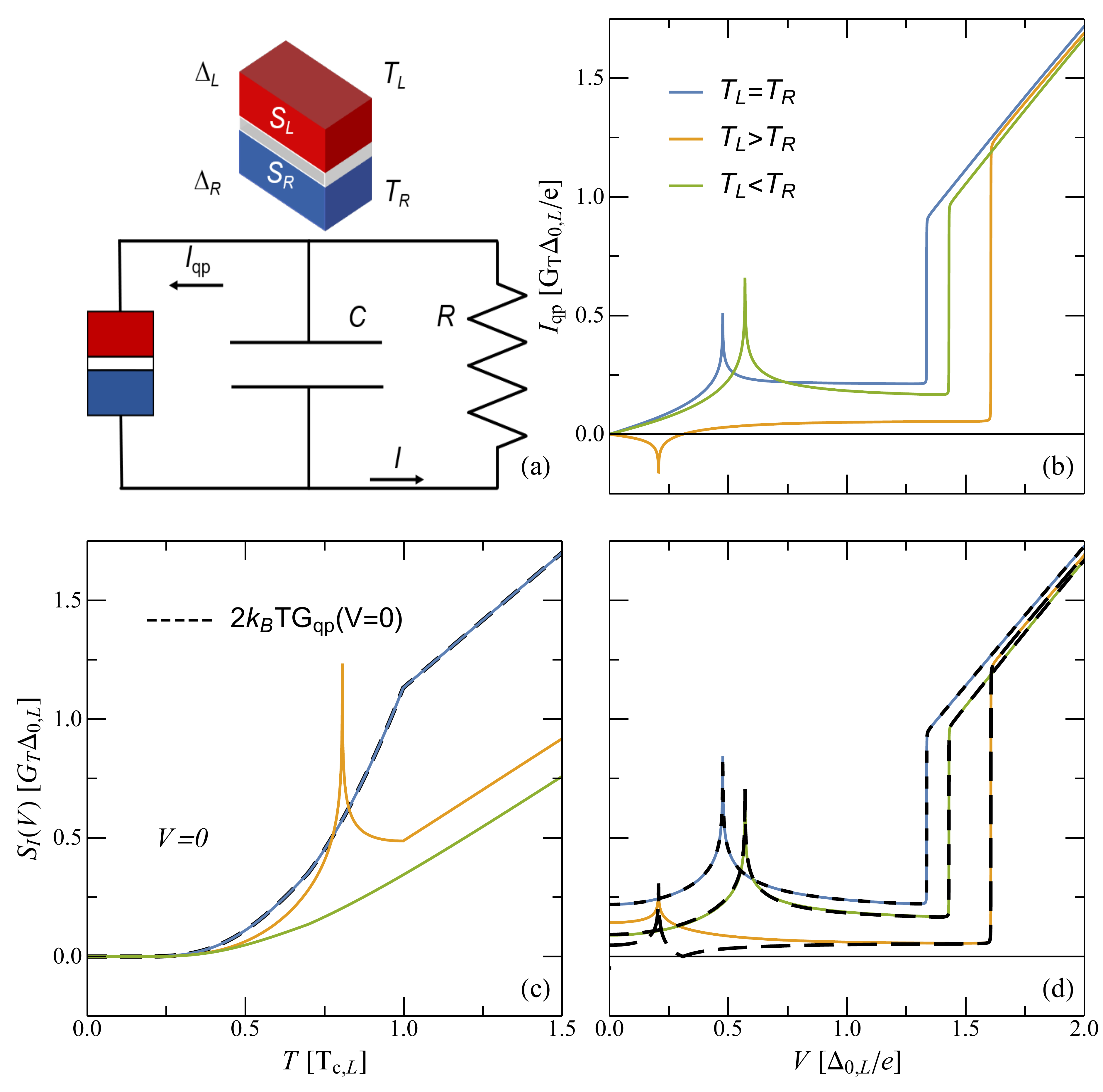}
		\caption{\textbf{(a)} Circuit scheme of the system. A thermally biased tunnel junction between two superconductors with different energy gap (SIS' junction) has capacitance $C$ and is connected to an external circuit with resistance $R$. \textbf{(b)} Quasiparticle current-voltage characteristic of a thermally biased tunnel junction between two BCS superconductors (SIS' junction) for $r=0.7$ and $T_{L}=T_{R}=0.6 T_{c,L}$ (blue), $T_{L}=0.6 T_{c,L}$, $T_{R}=0.01 T_{c,L}$ (orange), and $T_{R}=0.6 T_{c,L}$, $T_{L}=0.01 T_{c,L}$ (green). The thermoelectric behavior is characterized by a negative current for a positive voltage bias. \textbf{(c)} Temperature evolution of the zero-bias current noise $S_I(V=0)$ for $T=\max (T_L,T_R)$ and the different temperature gradients in panel (a). For $T_L\neq T_R$, the cold temperature is set to $T_{\rm cold}=\min(T_L,T_R)=0,01 T_{c,L}$. The dashed black curve displays the Johnson-Nyquist relation at the thermal equilibrium (see text for more details).  \textbf{(d)} Voltage evolution of the zero-bias current noise for the values used in the panel (a). Dashed curves represent the generalized nonequilibrium fluctuation-dissipation expression discussed in the main text.
		}
		\label{Fig1}
	\end{centering}
\end{figure}

We consider the charge transport in a tunnel junction between two Bardeen-Cooper-Schrieffer (BCS) superconducting electrodes $L,R$. For simplicity, we assume to be able to suppress completely the contribution associated with the Josephson effect, as already discussed in Refs.~\onlinecite{MarchegianiPRL,MarchegianiPRB}. However, we have recently observed that the presence of a moderate Josephson effect does not completely spoil the thermoelectric phenomenology~\cite{marchegiani2020phasetunable}. In this case, the transport is given by the quasiparticles only. For BCS superconductors, the quasiparticle density of states (smeared by the Dynes parameter $\Gamma$~\cite{Dynes1978,Dynes1984}) reads $N_j(E)=|\Re[(E+i\Gamma_j)/\sqrt{(E+i\Gamma_j)^2-\Delta_j^2}]|$ (with $j=L,R$), where $E$ is the quasiparticle energy with respect to the chemical potential (either $\mu_L$ or $\mu_R$) and $\Delta_j(T_j)$ are the temperature dependent superconducting order parameters, which are computed self-consistently~\cite{Tinkham2004}. Since we are interested in the description of thermoelectric phenomena, we assume that each of the two electrodes is in thermal equilibrium, namely the energy distribution follows the Fermi function $f_j(E)=[\exp(E/k_B T_j)+1]^{-1}$. Hereafter, we consider also situations where the two temperatures can be different $T_L\neq T_R$. The quasiparticle current through the junction in the tunneling regime reads~\cite{barone1982physics}
\begin{equation}
I_{\rm qp}=\frac{G_T}{e}\int_{-\infty}^{+\infty}N_L(E+eV)N_R(E)[f_R(E)-f_L(E+eV)]dE
\label{eq:iqp}
\end{equation}
where $-e$ is the electron charge, $G_T$ is the normal state conductance of the junction, and $V=(\mu_L-\mu_R)/(-e)$ is the voltage bias. In Refs.~\onlinecite{MarchegianiPRL,MarchegianiPRB}, we demonstrated that a junction between two superconductors with different gaps, i.e. for $\Delta_{ 0,L}\neq \Delta_{0,R}$ [ $\Delta_{ 0,j}=\Delta_j(T_j=0)=1.764 k_B T_{c,j}$, where $T_{c,j}$ is the critical temperature of the superconductor $j$], can support a nonlinear thermoelectric effect when the superconductor with the largest gap is heated up. For this reason, we introduce a symmetry parameter $r=\Delta_{0,R}/ \Delta_{0,L}$ and we set $r<1$ with no loss of generality. For this choice, the thermoelectric effect occurs only for $T_L\gtrsim T_R/r$ if $\Delta_L(T_L)>\Delta_R(T_R)$~\cite{MarchegianiPRB}. Figure~\ref{Fig1}(b) displays the voltage-current characteristic~\footnote{Only positive values of the voltage bias $V>0$ are considered, since $I(-V)=-I(V)$ due to the electron-hole symmetry.} for $r=0.7$ and different values of the thermal gradient across the junction. The evolution of the curves is non-monotonic for subgap voltages $V<|\Delta_L(T_L)+\Delta_R(T_R)|/e$, characterized by a peak at $V_p=|\Delta_L(T_L)-\Delta_R(T_R)|/e$ (due to the matching of the BCS singularities in the density of states~\cite{Tinkham2004,barone1982physics}), and monotonically increasing for $V>|\Delta_L(T_L)+\Delta_R(T_R)|/e$ [where the current for $eV\gg (\Delta_{L}+\Delta_{R})$ asymptotically reads $I_{\rm qp}\sim G_T V$)]. Moreover, the junction is always dissipative, i.e., $I_{\rm qp}V>0$, when the two electrodes are at the thermal equilibrium or for $T_R>T_L$~\footnote{Conversely, for $r>1$, the junction is always dissipative for $T_L\geq T_R$~\cite{MarchegianiPRL}}. Conversely, for $T_L>T_R$, we can have a thermoelectric generation of power, characterized by a negative current for a positive voltage bias (absolute negative conductance~\cite{MarchegianiPRL}). As a consequence, there is a finite value of the voltage bias $V= V_S\gtrsim V_p$ (and a symmetric value at $V=-V_S$, not shown here), that represents the Seebeck voltage, where the current is zero, i.e., $I(V_S)=0$.

As discussed above, in this work we wish to focus on the description of the noise contributions in the presence of a voltage and a temperature bias. In the tunneling limit considered here specifically, the statistic associated with the stochastic tunneling of quasiparticles is given by a bidirectional Poissonian~\cite{LevitovPRB70}, and all the cumulants of the distributions can be accordingly computed~\footnote{Notably, except for a prefactor associated to the charge carrier, all the odd moments are proportional to the current, while all the even moments are proportional to the noise.}. In particular, the spectral density of the current fluctuations can be computed as the sum of the two tunneling rates (from $L$ to $R$ and \textit{vice versa}), and reads~\cite{Golubevnoise}

\begin{align}
S_I(V)=G_T&\int_{-\infty}^{+\infty}N_L(E+eV)N_R(E)\{f_L(E+eV)[1-f_R(E)]\nonumber\\
&+f_R(E)[1-f_L(E+eV)]\}dE.
\end{align}
This quantity is always positive by definition and it is a even function of the voltage bias, due to the electron-hole symmetry of the density of states $N_j(E)=N_j(-E)$. It necessarily includes both the contribution due to the shot noise and the contribution of the thermal fluctuations (Johnson-Nyquist noise). Indeed, at the thermal equilibrium, where $T_L=T_R=T$, one recovers the standard nonequilibrium fluctuation-dissipation theorem which prescribes~\cite{LevitovPRB70,Rogovin}
\begin{equation}
S_I(V,T)=eI_{\rm qp}(V,T)\coth[eV/(2k_B T)].
\label{eq:fluctdiss}
\end{equation}
 In the limit $eV\ll k_B T$ (at equilibrium $T_L=T_R=T)$, one recovers the Johnson-Nyquist relation
$S_{I}(V=0,T)=2 k_B T \mathcal G_{\rm 0,qp}(T)$, where $\mathcal G_{\rm 0,qp}=dI_{\rm qp}/dV|_{V=0}$ is the zero-bias differential conductance of the quasiparticle current of Eq.~\eqref{eq:iqp}. This relation is displayed in Fig.~\ref{Fig1}(c), where the temperature evolution of $S_{I}(V=0)$ is shown for $T_L=T_R=T$ (solid blue), and the differential conductance (multiplied by $2k_B T$) is drawn with a dashed line. In particular, the noise increases monotonically with the temperature. Note that the noise is strongly suppressed for $T\lesssim 0.4 T_{c,L}$, due to the gaps in the density of states of the two superconductors, which suppress the differential conductance. Finally, when $T>T_{c,L}$ both the electrodes are in the normal state (since $T_{c,R}<T_{c,L}$ for $r<1$), and the noise is simply linear in the temperature, with $\mathcal G_{\rm 0,qp}(T>T_{c,L})=G_T$. In the same plot, we also display the temperature evolution of $S_{I}(V=0)$ in the non-equilibrium case, where a thermal bias is included, either for $T_L=T>0.01 T_{c,L}=T_R$ (orange) and for $T_R=T>0.01 T_{c,L}=T_L$ (green). Typically, the noise is smaller with respect to the thermal equilibrium case, where both the electrodes are hot, however, for $T_L>T_R$, the noise amplitude is non-monotonic for $T\sim 0.8 T_{c,L}$, due to the matching of the singularities in the superconducting density of states at zero voltage bias, and can be larger than the correspondent value at thermal equilibrium. In fact, we have $\Delta_L(0.8 T_{c,L})\sim 0.7\Delta_{0,L}=\Delta_R$, and correspondingly we have a peak in the temperature evolution of the thermal noise current. This situation cannot be met for $T_R>T_L$, since $\Delta_{0,R}<\Delta_{0,L}$ and the energy gap decreases monotonically by increasing $T_R$. As a consequence, the noise is symply monotonically increasing with $T_R$ for $T_R>T_L$.
\begin{figure*}[tp]
	\begin{centering}
		\includegraphics[width=\textwidth]{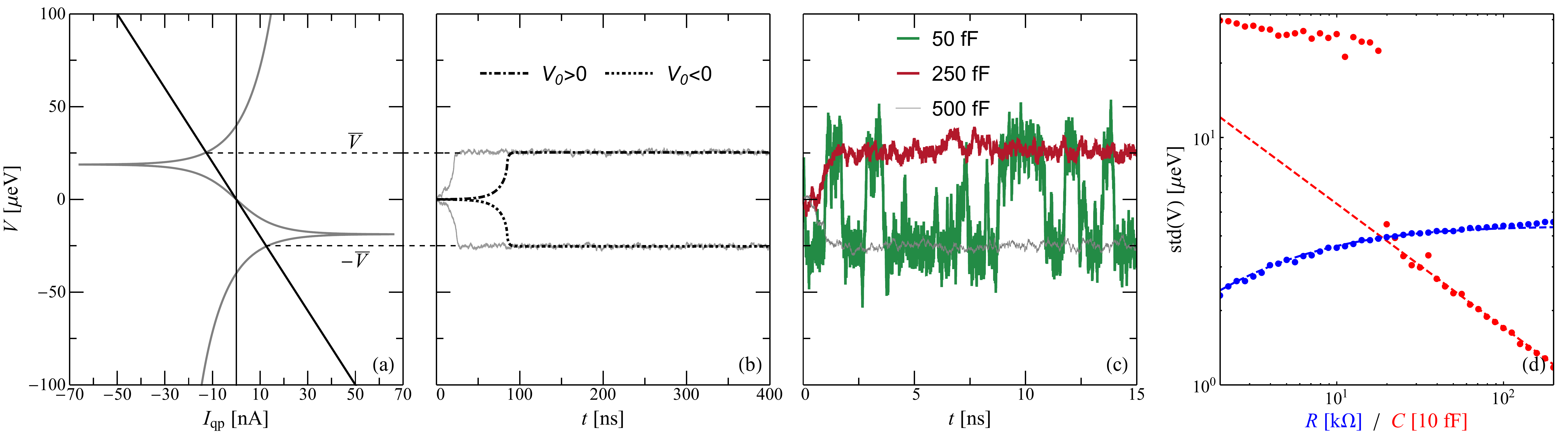}
		\caption{\textbf{(a)} Detail of the current-voltage characteristic of a thermally biased tunnel junction SIS' junction in the presence of thermoelectricity (gray). The black line is the load line of the resistor $R=2$k$\Omega$. The crossing points denote all the possible solutions of Eq.~\eqref{eq:stationary}, and represent the possible time independent steady-states. \textbf{(b)} Examples of the time dynamics of the junction either in the absence of noise (dashed) or in the presence of noise (solid) for the thermal bias of the panel (a). \textbf{(c)} Dynamics for different values of the capacitance of the junction, and thus of the noise current through it. \textbf{(d)} Standard deviation of the voltage fluctuation as a function of the capacitance and the load resistance of the system. The results of the numerical integration of Eq.~\eqref{eq:RCdynamics} (points) are compared with the approximate expression of Eq.~\eqref{eq:lineardVsq} (dashed). Parameters are $G_T=(1{\rm k}\Omega)^{-1}$, $r=0.75$, $T_{c,L}=1.6$K, $T_{L}=0.7 T_{c,L}$, and $T_{R}=0.01 T_{c,L}$.
		}
		\label{Fig2}
	\end{centering}
\end{figure*}

We consider now the full voltage evolution of $S_I(V)$ in Fig.~\ref{Fig1}(d) (solid curves), for the same values of the temperature of the two electrodes as in Fig.~\ref{Fig1}(b). The evolution is similar to the one displayed by the current, i.e., the evolution is non-monotonic at sub-gap voltages, displaying a peak for $V=V_p$, and it is almost linear for $V>|\Delta_L(T_L)+\Delta_R(T_R)|/e$. In particular, for $V\gg |\Delta_L(T_L)+\Delta_R(T_R)|/e$, the noise depends only slightly on the temperature bias, since it is almost proportional to the quasiparticle current, $S_I(V)\sim e I_{\rm qp}(V)$ (shot noise limit). For a comparison, in the plot we also display (dashed curves) a generalization of the nonequilibrium fluctuation-dissipation expression, i.e., $S_I(V)=eI_{\rm qp}(V,T_L,T_R)\coth[e V/(2k_B \bar T)]$~\footnote{For $T_L>T_R$, we take the absolute value of the expression, in order to avoid a negative value for the approximation for $|V|<V_S$.}, where we take the average temperature $\bar T=(T_L+T_R)/2$ (computed for the three different situations). This relation holds exactly at the thermal equilibrium $T_L=T_R$ (blue curve) and gives a very good approximation even in the presence of a thermal gradient for $T_R>T_L$, where no thermoelectric effect occurs. Conversely, in the presence of the the thermoelectric effect, i.e., $T_L>T_R$, the approximate expression is substantially inaccurate for subgap voltages $eV<\Delta_L+\Delta_R$. The deviation is quite striking around the Seebeck voltage, where the quasiparticle current is zero $I_{\rm qp}(V_S)=0$, while the noise contribution is necessarily finite, due to unavoidable thermal effects. This shows that in the presence of thermoelectricity the noise contribution must be explicitly computed and cannot be simply associated to the size of the charge current.

Now we wish to discuss how and at which degree the noise affects the thermoelectric effects first described in Ref.~\onlinecite{MarchegianiPRL}. In order to describe the dynamics of the junction, we consider the basic circuit schematically pictured in Fig~\ref{Fig1}(a). The junction's capacitance is modeled as a lumped element of capacitance $C$ and it is connected to an external load $R$. In this scheme, the dynamics of the junction is modeled by a first order nonlinear differential equation, which describes the current conservation in the circuit, in terms of the voltage bias $V$ across the junction, namely 
\begin{equation}
I=-\frac{V}{R}=C \dot V+I_{\rm qp}(V)+I_{n}(t).
\label{eq:RCdynamics}
\end{equation} 
More precisely, the total current in the circuit is the sum of the quasiparticle current $I_{\rm qp}$, the displacement current in the capacitance and the noise contribution $I_{n}$. Regarding the latter, we model it as a Markovian stochastic process with Gaussian distribution, characterized by the moments
\begin{align}
\langle I_{n}(t)\rangle=0,\quad \langle I_{n}(t)I_{n}(t')\rangle=S_I(V)\delta(t-t').
\label{eq:moments}
\end{align}
Our description is inspired by the standard treatment of noise in the resistively and capacitively shunted junction model (RCSJ model~\cite{barone1982physics,Tinkham2004}), but it includes the bias dependence of the noise amplitude in the thermoelectric case. In the absence of noise, we clearly recover the system investigated in Ref.~\onlinecite{MarchegianiPRL}. In particular, we demonstrated that, in the presence of a thermoelectric effect, i.e., $\mathcal G_{\rm 0,qp}<0$, the system develops a thermoelectric voltage if the load resistance is larger than a threshold value. In fact, for $R>1/|\mathcal G_{\rm 0,qp}|$ (and  $\mathcal G_{\rm 0,qp}<0$), the zero-current solution with $V=0$ is unstable and the voltage bias approaches a finite value limit, either $\pm\bar V$, which is a solution of the self-consistent
integral equation
\begin{equation}
I_{\rm qp}(\bar V)+\frac{\bar V}{R}=0.
\label{eq:stationary}
\end{equation}
Hereafter, we consider realistic parameters based on thin-film aluminum technology. Hence, we set $G_T=(1 k\Omega)^{-1}$, $T_{c,L}=1.6$ K (and thus $\Delta_{0,L}=1.764 k_B T_{c,L}\sim 240\mu$eV), $T_{c,R}=0.75 T_{c,L}\sim 1.2$ K. The two different values of the critical temperature of the two electrodes can be obtained, for instance, by varying the thickness of the aluminum thin film. We consider the typical situation in the presence of thermoelectricity displayed in Fig.~\ref{Fig2}a, where the detail of the current-voltage characteristic is displayed (gray). For convenience in the graphical visualization, we rotated the axes with respect to Fig.~\ref{Fig1}(b), showing the current in the abscissa and the voltage bias in the ordinate [see the discussion of Fig.~\ref{Fig2}(b) below]. The solutions of Eq.~\eqref{eq:stationary} are given by the crossings between the IV characteristic and the load curve (black) of equation $-V/R$ (in the plot, $R=2 \mathrm{k}\Omega>1/|\mathcal G_{\rm 0,qp}|\sim 1.15 \mathrm{k}\Omega$).
As discussed in Ref.~\onlinecite{MarchegianiPRL}, the zero-current solution is dynamically unstable and any voltage fluctuation, either externally applied or induced by noise fluctuations in the junction, leads the system in the thermoelectric state, with the electron-hole symmetry spontaneously broken $\bar V=-I_{\rm qp}(\bar V)R\sim \pm 25\mu$eV.
The instability of the zero-bias solution is investigated in Fig.~\ref{Fig2}(b), where the dynamics of the junction, corresponding to the thermal gradient of Fig.~\ref{Fig2}(a), is displayed assuming a non-zero value of the voltage at initial time $|V_0|\sim 0.2\mu$eV.

In the absence of noise, the voltage across the junction approaches deterministically the steady state value $\bar V=|\bar V|\mathrm{sign}(V_0)$ (dashed curves). In the presence of a small noise contribution (see the discussion below), the system evolves to a similar stationary limit, fluctuating  around the stationary solutions (either $\bar V$ or $-\bar V$). However, the specific solution (either $\bar V$ or $-\bar V$) may randomically change independently on the sign of $V_0$, especially if the initial value is small compared to the standard deviation of the noise fluctuations, i.e., $|V_0|\ll \sqrt{\langle\delta V^2\rangle}$. This is displayed in Fig.~\ref{Fig2}(b), where we demonstrate that both the steady state solutions can be obtained even for $V_0>0$ (solid curves).

As discussed above, the noise in the junction generates a fluctuation of the voltage signal across the tunnel junction. Since the thermoelectric effect is characterized by the presence of two stationary values, i.e., $\pm \bar V$, the noise can induce even a switching from 
$\bar V\rightarrow -\bar V$ and \textit{vice versa} when the fluctuations are of the order of the mean thermoelectric signal $\sqrt{\langle\delta V^2\rangle}\sim \bar V$. For this reason, it is important to limit the size of the fluctuations. To do that, we wish to estimate $\sqrt{\langle\delta V^2\rangle}$. We consider first a small noise contribution, where we expect a small perturbation of the thermoelectric solution in the steady-state. In this limit, it is convenient to rewrite the dynamics in terms of the displacement by the stationary solution $u=V-\bar V$, 
 \begin{equation}
C \dot u=-\frac{u +\bar V}{R} +I_{\rm qp}(\bar V+u)+I_{n}(t).
\end{equation}
Since we focus on a small perturbation around $\bar V$, we can consider a first order expansion of the quasiparticle current $I_{\rm qp}(\bar V+u)\simeq I_{\rm qp}(\bar V)+\mathcal G_{\rm qp}(\bar V) u$, where  $\mathcal G_{\rm qp}(\bar V)=dI_{\rm qp}/{dV}|_{\bar V}$. Finally, we approximate $S_I(V)\simeq S_I(\bar V)$ in the noise term $I_{n}(t)$. By exploiting Eq.~\eqref{eq:stationary}, we obtain a linearized Langevin equation~\cite{risken1996fokker,coffey2004langevin}
\begin{align}
\dot u=-\gamma(\bar V) u+\eta(t),
\label{eq:linearLangevin}
\end{align}
where we identify an effective friction $\gamma(\bar V)=\left[1/R+\mathcal G_{\rm qp}(\bar V)\right]/C$ and the fluctuating term $\eta(t)=I_{n}(t)/C$. Note that $\gamma(\bar V)$ represents a proper friction irrespectively of $R$, i.e., $\gamma(\bar V)$>0, since the differential conductance at the thermoelectric value $\bar V$ is always a positive number $\mathcal G_{\rm qp}(\bar V)>0$, being $V_p<\bar V<V_S$ [see Fig.~\ref{Fig1}(b) and Fig.~\ref{Fig2}(a)], and as required by the stability analysis~\cite{MarchegianiPRL}. According to Eq.~\eqref{eq:moments}, the variance of the fluctuating term reads
\begin{align}
\langle \eta(t)\eta(t')\rangle=q\delta(t-t'),
\end{align}
where $q=S_{I}(\bar V)/C^2$. 
In the stationary regime, the amplitude of the voltage oscillations reads~\cite{risken1996fokker,coffey2004langevin}
\begin{equation}
\mathrm{ std}(V)=\sqrt{\langle\delta V^2\rangle}=\sqrt{\frac{q}{2\gamma}}=\sqrt{\frac{S_{I}(\bar V)}{2C[1/R+\mathcal G_{\rm qp}(\bar V)]}}.
\label{eq:lineardVsq}
\end{equation}
This result is approximately valid if the voltage fluctuations are small compared to the stationary solution $\bar V$, that gives the inequality, $S_I(\bar V)\ll 2C\bar V^2[1/R+\mathcal G_{\rm qp}(\bar V)]$. The expression of Eq.~\eqref{eq:lineardVsq} gives a simplified view of the impact of the various parameters in the amplitude of the fluctuations. First, the capacitance of the junction plays a crucial role, since it is
inversely proportional to the noise bandwidth. Conversely, the role of the other parameters is more complex. In fact, the value of the stationary solution $\bar V$ depends on the ratio between the load resistance $R$ and the tunnel conductance of the junction $G_T$. In typical operating conditions, where it is required $R>1/|\mathcal G_{\rm 0,qp}|$, the term $1/R$ can be neglected in the square bracket when $\mathcal G_{\rm qp}(\bar V)R\gg 1$.
In this limit, the voltage fluctuation reads ${\rm std}(V)\sim\sqrt{S_I(\bar V)/[2C \mathcal G_{\rm qp}(\bar V)]}$. For this case, since  $S_I, \mathcal G_{\rm qp}\propto G_T$, the variance of the signal is independent on the normal state conductance of the junction $G_T$ and it is only affected by the electrical capacitance $C$.
 
To test the predictions of the perturbative approach, we solved numerically Eq.~\eqref{eq:RCdynamics} for some values of the capacitance of the junction, keeping fixed the load resistor $R=2{\rm k}\Omega$. The results are shown in Fig.~\ref{Fig2}(c), where the time dynamics is displayed for different values of $C$. Since the value of $R$ is the same as in Fig.~\ref{Fig2}(b), the stationary states are still given by $\bar V=\pm 25\mu$eV. As discussed above, the amplitude of the voltage fluctuations is minimized for the largest value of $C$ (in the figure $C=500$fF), and increases by lowering $C$. When the noise contribution is too strong (for $C=50$fF in Fig.~\ref{Fig2}), the fluctuations induce a stochastic switching between the two thermoelectric solutions. In order to quantify the degree of validity of our approximation Eq.~\eqref{eq:lineardVsq}, we evaluate numerically the standard deviation of the voltage signal in the steady state as a function of the capacitance of the junction, displayed in Fig.~\ref{Fig2}d (red points). For $C\gtrsim 150$ fF, ${\rm std}(V)\propto 1/\sqrt{C}$, in full agreement with the theoretical expression Eq.~\eqref{eq:lineardVsq}, displayed with a dashed line (and computed for $\bar V=25\mu$eV and $R=2$k$\Omega$). At lower values of $C$, the simplified expression becomes inaccurate since the small fluctuation approximation is no longer valid and, in particular, the noise can induce a switching between the two thermoelectric states, with a zero average value of the thermoelectric signal. In the same plot, we display also the dependence of ${\rm std}(V)$ on the load resistor $R$ (blue points), obtained by solving Eq.~\eqref{eq:RCdynamics} for $C= 0.5$ pF and different values of $R>2$k$\Omega$. The noise amplitude slightly increases monotonically with the load, and it is well described (blue dashed) by the linearized expression Eq.~\eqref{eq:lineardVsq} for the chosen value of the capacitance.

In summary, we computed the strength of the electron noise in a thermally biases superconducting tunnel junction, in a nonequilibrium situation which goes beyond the regime where the fluctuation-dissipation theorem usually holds. By solving the equation of motion of the voltage signal across the junction, we demonstrated the robustness to the noise of the nonlinear thermoelectric effect recently predicted in Ref.~\onlinecite{MarchegianiPRL}. While large voltage fluctuations would potentially hide the thermoelectric signal, the impact of the noise can be opportunely reduced by tuning the circuit parameters. More precisely, we demonstrated that the effective noise can be lowered by increasing the capacitance of the junction. This can be realized, for instance, by increasing the size of the junction (for an aluminum structure, where $C/A=50$fF$/\mu$m$^2$, a junction area of $A=3\times 3\mu{\rm m}^2$ would be sufficient to have a clear observation of the effect), or using an external electrical capacitor in parallel. Our results represent an important step in the experimental demonstration of the nonlinear thermoelectric effect in controlled configurations, and are relevant in the context of thermal management in superconducting quantum circuits where noise effects are not well explored.

The authors acknowledge the EU's Horizon 2020 research and innovation programme under grant agreement No. 800923 (SUPERTED) for
partial financial support. A.B. acknowledges the CNR-CONICET
cooperation program "Energy conversion in quantum nanoscale hybrid devices", the SNS-WIS joint lab QUANTRA funded by the Italian Ministry of Foreign Affairs and International Cooperation, and the Royal Society through the International Exchanges between the UK and Italy (Grants No. IES R3 170054 and IEC R2 192166).

\end{document}